\documentclass[12pt]{article}
\def\bseq{\begin{subequation}}  % = 1a 1b
\def\eseq{\end{subequation}}
\def\bsea{\begin{subeqnarray}}  % = 1.1a 1.1b
\def\esea{\end{subeqnarray}}
\newcommand{\bbox}{\lower.2ex\hbox{$\Box$}}

\newcommand{\beq}{\begin{equation}}
\newcommand{\eeq}{\end{equation}}
\newcommand{\bea}{\begin{eqnarray}}
\newcommand{\eea}{\end{eqnarray}}
\newcommand{\ena}{\end{eqnarray}}

\renewcommand{\a}{\alpha}
\renewcommand{\b}{\beta}

\renewcommand{\d}{\delta}

\newcommand{\pa}{\partial}

\newcommand{\G}{\Gamma}

\newcommand{\e}{\epsilon}

\renewcommand{\l}{\lambda}
\renewcommand{\L}{\Lambda}
\newcommand{\m}{\mu}

\newcommand{\n}{\nu}

\newcommand{\p}{\pi}

\newcommand{\s}{\sigma}

\newcommand{\Db}{\bar{D}}

\newcommand{\ad}{{\dot{\alpha}}}
\newcommand{\bd}{{\dot{\beta}}}

\begin{document}
\begin{titlepage}
\begin{flushright}
IFUM-FT-662 \\

%hep-th/0009196
\end{flushright}
\vspace{1cm}
\begin{center}
{\Large \bf Noncommutative perturbation in superspace}
\vfill%\vskip 15mm%27.mm
{\large \bf
Daniela Zanon}\\
%\vfill
\vskip 7mm%1cm
{\small
Dipartimento di Fisica dell'Universit\`a di Milano
and\\ INFN, Sezione di Milano, Via Celoria 16,
20133 Milano, Italy\\}
\end{center}
\vfill
\begin{center}
{\bf Abstract}
\end{center}
{\small We consider noncommutative ${\cal N}=4$ supersymmetric
$U(N)$ Yang-Mills theory. Using the ${\cal N}=1$ superfield formalism
and the background field method we compute  one-loop four point
contributions to the effective action and compare the result with the
field theory limit from open string amplitudes in the presence of a
constant $B$-field.}
\vspace{2mm}
\vfill
\hrule width 3.cm
\begin{flushleft}
e-mail: daniela.zanon@mi.infn.it
\end{flushleft}
\end{titlepage}

String theory dynamics in the presence on a nonzero $B$-field leads
to the appearance of noncommutative geometry, giving rise in the
field theory limit to noncommutative gauge theories \cite{all}.

Field theories on a noncommutative spacetime can be formulated using standard
field theoretical methods, trading the noncommutative property of the geometry
with a deformation of the multiplication rule between the fields. In
practice one simply replaces the ordinary product between the
fields with the $*$-product, which absorbs and contains the non local nature
of the noncommutative theory. In this framework one can proceed and
quantize the theory: the Feynman rules are  modified by the appearance of
exponential factors at the vertices, but otherwise standard
perturbation theory can be applied. Indeed a lot of progress has been
made in this direction \cite{filk,perturb}.

Supersymmetric versions of noncommutative field theories have also
been considered and in particular their formulation in superspace has
been presented \cite{FL,susy}. In this letter we address the issue of
quantization and perturbation in a superfield-superspace approach.
The supersymmetric noncommutative theory is defined on standard
superspace, while superfields are multiplied via the $*$-product.
The $*$-operation does not touch the fermionic
coordinates and simply introduces derivatives of superfields which
are themselves superfields again.
Therefore the quantization is performed in a
standard manner. The only modifications are in the interaction
terms which contain exponential factors from the $*$-product.
One constructs supergraphs and performs the
$D$-algebra in the loops with no new rules as compared to the
commutative case. Once this is done, one is left with momentum
integrals which are of the same kind as for bosonic noncommutative
theories.

As an illustration of the general procedure outlined above,
we study the quantization of noncommutative ${\cal N}=4$ supersymmetric
$U(N)$ Yang-Mills theory in a ${\cal N}=1$ superfield setting.
We compute  one-loop four point
contributions to the effective action using the background field
method.
The result we obtain is in perfect agreement with the
field theory limit from open string amplitudes in the presence of a
constant $B$-field \cite{*trek}.

\vspace{0.8cm}

Using the definition of the $*$-product for superfields
\beq
(\phi_1 * \phi_2)(x,\theta,\bar{\theta})\equiv e^{\frac{i}{2}
\Theta^{\m\n}\frac{\pa}{\pa x^\m}\frac{\pa}{\pa y^\n}}~
\phi_1(x,\theta,\bar{\theta})\phi_2(y,\theta,\bar{\theta})|_{y=x}
\label{starprod}
\eeq
the classical action for the noncommutative
${\cal N}=4$ supersymmetric Yang-Mills theory,
in terms of ${\cal N}=1$ superfields, can be written as (we use the
notations and conventions adopted in \cite{superspace})
\bea
&&S= \frac{1}{g^2}~{\rm Tr} \left( \int~ d^4x~d^4\theta~ e^{-V}
\bar{\Phi}_i e^{V} \Phi^i +\frac{1}{2} \int ~d^4x~d^2\theta~ W^2
+\frac{1}{2} \int ~d^4x~d^2\bar{\theta}~ \bar{W}^2 \right.\nonumber\\
&&\left.\left.~~~~~~~~~~~~~+\frac{1}{3!} \int ~d^4x~d^2\theta~ i\e_{ijk}
 \Phi^i
[\Phi^j,\Phi^k] + \frac{1}{3!}\int ~d^4x~d^2\bar{\theta}~ i\e^{ijk} \bar{\Phi}_i
[\bar{\Phi}_j,\bar{\Phi}_k] \right)\right|_*
\label{SYMaction}
\eea
where the symbol $|_*$ implies that the multiplication of superfields is
performed as defined in (\ref{starprod}).
In (\ref{SYMaction}) the $\Phi^i$ with $i=1,2,3$ denote three chiral
superfields, while $W^\a= i\bar{D}^2(e^{-V}D^\a e^V)$ is the gauge
superfield strength. All the fields are Lie-algebra valued, e.g.
$\Phi^i=\Phi^i_a T_a$, and $T_a$ are $U(N)$ matrices.
The $*$-product in (\ref{starprod}) not only maintains  explicit
supersymmetry; it also preserves gauge invariance. In fact it is easy
to show that the action in (\ref{SYMaction})
is invariant under nonlinear gauge transformations,
which are just the obvious generalization to the noncommutative case
of the standard ones \cite{FL,superspace}
\bea
e^V&\rightarrow& e_*^{i\bar{\L}}*e_*^V*e_*^{-i\L}\nonumber\\
&&~~~~~~\nonumber\\
\Phi~\rightarrow~
e_*^{i\L}*\Phi* e_*^{-i\L}\qquad&&\qquad\bar{\Phi}~\rightarrow~
e_*^{i\bar{\L}}*\bar{\Phi}*e_*^{-i\bar{\L}}
\label{gaugetransf}
\eea
with a gauge parameter $\L$ which is a chiral superfield.

Now we want to study one-loop corrections to the effective action. We
perform the quantization directly in superspace and take advantage of
the background field method which greatly simplifies the
calculations. For ordinary commutative theory it has the additional,
main property of keeping explicit the gauge invariance of the result,
at every stage of the perturbative computation. For the noncommutative theory
defined in (\ref{SYMaction}) we will find that the
one-loop effective action is still expressible in terms of field strengths, as
the background field method guarantees, but the $*$-product is not
maintained. This result confirms what expected from one-loop
string theory \cite{*trek,gar}.

The background field quantization  has been used efficiently in perturbative
calculations for commutative SYM theory
\cite{GS,superspace,GZ}. We briefly summarize the method and the results in
order to extend them to the noncommutative case.

The quantum-background splitting can be most easily
formulated in terms of covariant derivatives. To this end  first
one rewrites the gauge Lagrangian as
\beq
{\rm Tr}~ W^2= -{\rm Tr} \left( \frac{1}{2}[\bar{\nabla}^\ad,\{\bar{\nabla}_\ad,
\nabla_\a\}]\right)^2
\label{lagrang}
\eeq
with
\beq
\nabla_\a =e^{-\frac{V}{2}}~ D_\a~ e^{\frac{V}{2}}\qquad
\qquad \bar{\nabla}_\ad=e^{\frac{V}{2}}~ \bar{D}_\ad ~e^{-\frac{V}{2}}
\label{covder}
\eeq
Then one performs the splitting by rewriting them
in terms of the quantum prepotential $V$ and background covariant
derivatives
\beq
\nabla_\a ~\rightarrow~e^{-\frac{V}{2}}~\nabla_\a ~ e^{\frac{V}{2}}\qquad
\qquad \bar{\nabla}_\ad~\rightarrow~
e^{\frac{V}{2}}~ \bar{\nabla}_\ad ~e^{-\frac{V}{2}}
\label{backcovder}
\eeq
where now the covariant derivatives are expressed in terms of
background connections, i.e.
\beq
\nabla_\a= D_\a-i{\bf{\G}}_\a \qquad \qquad \bar{\nabla}_\ad= \bar{D}_\ad-
i\bar{\bf{\G}}_\ad \qquad\qquad \nabla_a=\pa_a-i{\bf{\G}}_a
\label{backcovderconn}
\eeq
The quantum gauge invariance is fixed through the introduction of
background covariantly
chiral gauge fixing functions, $\nabla^2 V$ and $\bar{\nabla}^2 V$.
 When added to the classical Lagrangian
(\ref{lagrang}) they lead to
\beq
-\frac{1}{2g^2} {\rm Tr}\left[ \left(e^{-V}\nabla^\a e^V\right)
\bar{\nabla}^2  \left(e^{-V}\nabla_\a e^V\right)
+ V\left(\nabla^2\bar{\nabla}^2+\bar{\nabla}^2
\nabla^2\right)V\right]
\eeq
so that the quantum quadratic gauge Lagrangian can be written as
\beq
-\frac{1}{2g^2} {\rm Tr}~ V~\left[\frac{1}{2}\nabla^a\nabla_a
 -i {\bf W}^\a\nabla_\a
-i \bar{{\bf W}}^\ad\bar{\nabla}_\ad \right] V
\label{quadratic}
\eeq
where $\frac{1}{2}\nabla^a\nabla_a $
is the background covariant d'Alembertian and ${\bf W}^\a$ is the
background field strength. Since we are interested in one-loop
calculations we only need terms in the action which are quadratic
in the quantum fields. Thus the expression in (\ref{quadratic})
suffices: from there one can isolate a free kinetic term plus
interactions with the background. Using the definitions of the connections
in (\ref{backcovderconn}) one finally obtains
\bea
&&-\frac{1}{2g^2} {\rm Tr} ~V\left[\frac{1}{2} \pa^a\pa_a-i{\bf \G}^a\pa_a
-\frac{i}{2} \pa^a{\bf\G}_a -\frac{1}{2}{\bf\G}^a{\bf
\G}_a \right.\nonumber\\
&&\left.~~~~~~~~~~~~~~~ -i {\bf W}^\a (D_\a-i{\bf\G}_\a)-i \bar{{\bf W}}^\ad
( \bar{D}_\ad-i\bar{{\bf \G}}_\ad)\right]V
\label{oneloopaction}
\eea
The quantum vector fields have standard propagators and interactions
with the background that one reads from (\ref{oneloopaction}).

The gauge-fixing procedure requires the introduction of
ghost fields \cite{superspace}. We have two Faddeev-Popov ghosts
$c$ and $c'$. Moreover, since we
have chosen background-covariantly chiral gauge-fixing
functions, we need a Nielsen-Kallosh ghost $b$. They are  all background
covariantly chiral
superfields, i.e. $\bar{\nabla}_\ad c=\bar{\nabla}_\ad c'=
\bar{\nabla}_\ad b= 0$.

For ${\cal N}=4$ supersymmetric Yang-Mills we have in addition the three
background covariantly chiral matter superfield $\Phi^i$
(~see (\ref{SYMaction})~).
In this case one-loop contributions to the effective action
with external vector
fields can be easily computed \cite{GS2,superspace}. One finds that for
a diagram with an
arbitrary number of external vector background lines, the loops from the three
chiral matter fields are exactly cancelled
by the corresponding loops from the three chiral ghosts which have opposite
statistics.
Only quantum vector
loops survive and they give the first nonvanishing result at the
level of the four-point function. This is due to the fact that
superspace Feynman rules require the presence of two $D$'s and
two $\bar{D}$'s for a non
zero loop contribution.
From the action in (\ref{oneloopaction}) we have interactions with
the background fields at most linear in the $D$'s. Therefore at least
four vertices are needed.
The calculation is straightforward and leads to a very simple result:
the four-point vector amplitude is given by
\cite{superspace}
\bea
&&\G= \frac{1}{2} {\rm Tr}~ \int d^2\theta~d^2\bar{\theta}~
\frac{d^4p_1~d^4p_2~d^4p_3~d^4p_4}{(2\p)^{16}}~\d(\sum
p_i)~G_0(p_1\dots p_4)\nonumber\\
&&~~~~~~~~~\left[{\bf{W}}^\a(p_1){\bf{W}}_\a(p_2) \bar{{\bf{W}}}^\ad(p_3)
\bar{{\bf{W}}}_\ad(p_4)-\frac{1}{2}
{\bf{W}}^\a(p_1)\bar{{\bf{W}}}^\ad(p_2){\bf{W}}_\a(p_3)\bar{{\bf{W}}}_\ad(p_4)
\right]\nonumber\\
&&~~~
\label{fourpoint}
\eea
where $G_0$ is the four point scalar integral
\beq
G_0=\int d^4k~ \frac{1}{(k+p_1)^2 k^2 (k-p_4)^2 (k+p_1+p_2)^2}
\label{scalarbox}
\eeq
In order to make contact with corresponding calculations in ordinary
Yang-Mills theory, we observe that from (\ref{fourpoint}) we obtain the expected
bosonic expression, i.e.
\bea
&&\int~d^2\theta~d^2\bar{\theta}\left[{\bf{W}}_a^\a(p_1){\bf{W}}_{\a b}(p_2)
 \bar{{\bf{W}}}_c^\ad(p_3)
\bar{{\bf{W}}}_{\ad d}(p_4)\right.\nonumber\\
&&\left.~~~~~~~~~~~~~~-\frac{1}{2}
{\bf{W}}_a^\a(p_1)\bar{{\bf{W}}_b}^\ad(p_2){\bf{W}}_{\a c}(p_3
)\bar{{\bf{W}}}_{\ad d}(p_4)
\right]\nonumber\\
&&~~~~\nonumber\\
&&\rightarrow
\frac{1}{4}\left[ F^{\m\n}_a F_{\n\s b}F_{\m\rho c}F^{\rho\s}_d
+\frac{1}{2}F^{\m\n}_a F_{\n\s b}F^{\rho\s}_c F_{\m\rho
d}\right.\nonumber\\
&&~~~~~~~~~~~~~\left.-\frac{1}{4}F^{\m\n}_aF_{\m\n b}F^{\rho\s}_cF_{\rho\s d}
-\frac{1}{8}F^{\m\n}_aF^{\rho\s}_bF_{\m\n c}F_{\rho\s d}\right]
\label{bos}
\eea
Now we want to repeat the computation for the
{\em noncommutative} theory.

\vspace{1.5cm}

We go back to the ${\cal N}=4$ Yang-Mills action
in (\ref{SYMaction}). Since, as already emphasized, the $*$-product
does not affect superspace properties, it is clear that the various steps
of the background field quantization can be implemented even in this
case.  We can go all the way to the action in  (\ref{oneloopaction})
and there too we replace the ordinary
multiplication between superfields with the $*$-operation. Following
what we have
done in the commutative example,  we
consider terms in the effective action with external vector fields.
At the
one-loop level again we find that ghost contributions cancel matter
superfield contributions and one has to deal only with vector quantum
loops. As before, for $D$-algebra reasons, the two- and three-point
functions are zero and the first nonvanishing result is a loop with
four vertices. Thus we focus on this calculation.

\vspace{0.8cm}

We have to compute a box supergraph with Feynman rules
that can be obtained directly from (\ref{oneloopaction}) with the appropriate
$*$-multiplication inserted. For any noncommutative theory the
quadratic part of the action is the same as in the commutative case.
Thus we have in momentum space
the vector propagators given by
\beq
<V^a(\theta)V^b(\theta')>=-\frac{g^2}{p^2}\d^{ab} \d^4(\theta-\theta')
\label{prop}
\eeq
The relevant vertices are (~cf. (\ref{oneloopaction})~)
\beq
\frac{1}{2g^2} {\rm Tr} ~V*\left[~i {\bf W}^\a D_\a+i \bar{{\bf W}}^\ad
\bar{D}_\ad~\right]*V
\label{vertices}
\eeq
and one needs two $D$'s and two $\bar{D}$'s for a nonzero completion
of the $D$-algebra in the loop.
As in (\ref{fourpoint}) we always obtain two ${\bf W}$'s and
two $\bar{{\bf W}}$'s, a factor  with four scalar
propagators, and in addition exponential factors from
the $*$-product
at the vertices.

More precisely, using the definition in (\ref{starprod}), the
three-point interactions can be written in momentum space as
\bea
&&\frac{1}{2g^2}~\left({\cal U}(k_1,k_2,k_3) +\bar{\cal U}(k_1,k_2,k_3)
\right)\equiv\nonumber\\
&&~~~~~~~~~~~\equiv \frac{1}{2g^2}V_a(k_1)\left[~i {\bf W}_b^\a(k_2) D_\a+i
\bar{{\bf W}}_b^\ad(k_2)
\bar{D}_\ad~\right] V_c(k_3)~\nonumber\\
&&~~~~~~~~~~~~~~~~~~~~~~~~~~~~~~~~{\rm Tr}(T^aT^bT^c)~ e^{-\frac{i}{2}(k_1
\times k_2+k_2\times k_3+k_1\times k_3)}
\label{vertex}
\eea
with momenta flowing into the vertex,
$k_1+k_2+k_3=0$ and $k_i\times k_j\equiv
(k_i)_\m\Theta^{\m\n}(k_j)_\n$. In order to obtain a box diagram we
need a forth order term from the Wick expansion, with two ${\cal U}$'s
and two $\bar{\cal U}$'s, i.e. $\frac{1}{4(2g^2)^4}~{\cal U}^2~\bar{\cal U}^2$.
Every vertex, when inserted in the loop, gives rise to an
{\em untwisted} and a {\em twisted} term
\bea
&&{\cal U}(k_1,k_2,k_3)\rightarrow  V_a(k_1)~i {\bf W}_b^\a(k_2) D_\a
V_c(k_3)~\left[~{\rm Tr}(T^aT^bT^c)~ e^{-\frac{i}{2}(k_1
\times k_2+k_2\times k_3+k_1\times k_3)}\right.\nonumber\\
&&~~~~~~~~~~~~~~~~~~~~~~~~~~~~~~\left.-{\rm Tr}(T^cT^bT^a)~  e^{\frac{i}{2}(k_1
\times k_2+k_2\times k_3+k_1\times k_3)}\right]
\label{twistvert}
\eea
with the understanding that now the quantum lines have  to be Wick
contracted in the order in which they appear.
As in the commutative case, there are two possible
arrangements of the vertices in the loop,
i.e.  ${\cal U}(1){\cal U}(2)\bar{\cal U}(3)
\bar{\cal U}(4)$ with multiplicity two,
and ${\cal U}(1)\bar{\cal U}(2)
{\cal U}(3)\bar{\cal U}(4)$ with multiplicity one.

The $D$-algebra is trivial and it gives $D_\a D_\b \Db_\ad \Db_\bd\rightarrow
C_{\b\a} C_{\bd\ad}$ and $D_\a \Db_\ad D_\b  \Db_\bd\rightarrow
-C_{\b\a} C_{\bd\ad}$ respectively for the two arrangements of the
vertices.
We obtain a result that we can write symbolically as
\beq
\frac{1}{32}\left[~{\cal U}^\a~{\cal U}_\a~ \bar{\cal U}^\ad~ \bar{\cal
U}_\ad~ -~\frac{1}{2}~{\cal U}^\a~ \bar{\cal U}^\ad~ {\cal U}_\a~
\bar{\cal U}_\ad~\right]
\label{box}
\eeq
where we have defined (~see (\ref{twistvert})~)
\bea
&&{\cal U}^\a(k_1,k_2,k_3)
\equiv {\cal U}_P^\a(k_1,k_2,k_3)+{\cal U}_T^\a(k_1,k_2,k_3)
\equiv
V_a(k_1)~i {\bf W}_b^\a(k_2)
V_c(k_3)~\nonumber\\
&&~~~~~\nonumber\\
&&~~~~~~~~\left[~{\rm Tr}(T^aT^bT^c)~ e^{-\frac{i}{2}(k_1
\times k_2+k_2\times k_3+k_1\times k_3)}-{\rm Tr}(T^cT^bT^a)~  e^{\frac{i}{2}(k_1
\times k_2+k_2\times k_3+k_1\times k_3)}\right]\nonumber\\
&&~~~~~
\label{newvertex}
\eea
The $V$
quantum lines must be contracted in the consecutive order as they appear in
(\ref{box}). Substituting (\ref{newvertex}) in (\ref{box}), we obtain
the sum of sixteen terms: two of them, i.e. the ones which contain all
untwisted $P $ and all twisted $T $ vertices correspond to planar diagrams.
All the others, i.e. the ones with two $P$ and two $T$ vertices (a
total of six), the ones with one $P$ and three $T$'s (a total of four)
and the ones with one $T$ and three $P$'s (a total of four), correspond
to nonplanar graphs. Now we analyze these contributions in some
detail.

\vspace{0.8cm}

As anticipated above, the planar diagrams correspond to terms from
(\ref{box}) which
contain either four untwisted vertices ${\cal U}_P$ or four twisted
vertices ${\cal U}_T$. For the $U(N)$ gauge matrices we use the relation
$T^a_{ij}T^a_{kl}=\d_{il}\d_{jk}$, and
obtain the following contribution to the effective action
\bea
&&\G_{{\rm planar}}= \frac{N}{32} ~ \int d^2\theta~d^2\bar{\theta}~
\frac{d^4p_1~d^4p_2~d^4p_3~d^4p_4}{(2\p)^{16}}~\d(\sum
p_i)~G_0(p_1\dots p_4)\nonumber\\
&&~~~~~\nonumber\\
&&~~~~~~~~~
\left( {\bf W}_a^\a(p_1){\bf W}_{\a b}(p_2) \bar{{\bf W}_c}^\ad(p_3) \bar{{\bf
W}}_{\ad d}(p_4)-\frac{1}{2}
{\bf W}_a^\a(p_1)\bar{{\bf W}}_b^\ad(p_2){\bf W}_{\a c}(p_3)\bar{{\bf
W}}_{\ad d}(p_4)\right)\nonumber\\
&&~~~~\nonumber\\
&&~~~~~~~~~~~~~{\rm Tr}(T^aT^bT^cT^d)~\left[ e^{\frac{i}{2}(p_2
\times p_1+p_1\times p_4+p_2\times p_4)}+e^{-\frac{i}{2}(p_2
\times p_1+p_1\times p_4+p_2\times p_4)}\right]
\label{planar}
\eea
with $G_0$ defined in (\ref{scalarbox}). Comparing (\ref{planar})
with the result in the commutative theory we find that  the
only difference is given by the exponential factors which depend
on $\Theta$  and on the external momenta \cite{filk}.
In fact the exponentials are such to reconstruct the $*$-product
between the field strengths, yielding
\bea
&&\G_{{\rm planar}}=\frac{N}{32} {\rm Tr}~ \int d^2\theta~d^2\bar{\theta}~
\frac{d^4p_1~d^4p_2~d^4p_3~d^4p_4}{(2\p)^{16}}~\d(\sum
p_i)~G_0(p_1\dots p_4)~~~~~~~~~~~~~~~~~~\nonumber\\
&&~~~~\nonumber\\
&&~~
\left( {\bf W}^\a(p_1)*{\bf W}_{\a }(p_2)*
 \bar{{\bf W}}^\ad(p_3)* \bar{{\bf W}}_{\ad }(p_4)+
\bar{{\bf W}}^{\ad }(p_4)*\bar{{\bf W}}_\ad(p_3)*{\bf W}^{\a }(p_2)*
{\bf W}_\a(p_1)
\right.\nonumber\\
&&\left.-\frac{1}{2}
{\bf W}^\a(p_1)*\bar{{\bf W}}^\ad(p_2)*{\bf W}_{\a }(p_3)*
\bar{{\bf W}}_{\ad }(p_4)-\frac{1}{2}\bar{{\bf W}}^{\ad }(p_4)*
{\bf W}^{\a }(p_3)*\bar{{\bf W}}_\ad(p_2)*{\bf W}_\a(p_1)
\right)\nonumber\\
&&~~~
\label{boxplanar}
\eea
Making use of the relation in (\ref{bos}) with appropriate $*$-products
implemented, we can obtain the bosonic expression corresponding to
(\ref{boxplanar}).
Now we turn to the study of the nonplanar supergraphs.

\vspace{0.6cm}

The various nonplanar
diagrams can be collected in two distinct groups. There are graphs
in which  two vertices are twisted. In this first class the $U(N)$ matrices
produce a factor like ${\rm Tr}(T^p T^q){\rm Tr}(T^rT^s)$.
Then there are the graphs in which one (or equivalently three) of the
four vertices are twisted. For them
the trace on the gauge matrices factorizes as ${\rm Tr}( T^p) {\rm Tr}(T^q
T^rT^s)$.
In both cases the phases from the $*$-product at the
vertices will contain a dependence on the loop momentum.

We illustrate the procedure
considering nonplanar graphs in the first  group, from a structure of
the vertices like $P P T T$.  Looking at (\ref{box}) we find that these
contributions come from
\beq
{\cal U}_P^\a~{\cal U}_{P\a}~ \bar{\cal U}_T^\ad~ \bar{\cal
U}_{T\ad} -~\frac{1}{2}~{\cal U}_P^\a~ \bar{\cal U}_P^\ad~{\cal U}_{T\a}~
\bar{\cal U}_{T\ad}
\label{WWbar}
\eeq
With the external momenta
in the order $p_1$, $p_2$, $p_3$, $p_4$, the exponentials from the
$*$-operation at the vertices give
\beq
e^{-\frac{i}{2}( -p_1 \times k)}~e^{-\frac{i}{2}[ -p_2 \times
(k+p_1)]}~
e^{\frac{i}{2}[(p_1+p_2+p_4)\times(k-p_4)]}~
e^{\frac{i}{2}(k \times p_4)}
\eeq
We find a  contribution to the effective action of the form
\bea
&&\G_{PPTT}=\frac{1}{32}~\int d^2\theta~d^2\bar{\theta}~
\frac{d^4p_1~d^4p_2~d^4p_3~d^4p_4}{(2\p)^{16}}~\d(\sum
p_i)~\nonumber\\
&&~~~~\nonumber\\
&&~~~~~~~~\int d^4k~
\frac{e^{i(p_1+p_2)\times k}}{(k+p_1)^2k^2(k-p_4)^2 (k+p_1+p_2)^2 }\nonumber\\
&&~~~~~~~~~~\nonumber\\
&&~~~~~~~~~~~e^{-\frac{i}{2}p_1\times p_2}~e^{\frac{i}{2}p_3\times
p_4}\left[{\rm Tr}({\bf W}^\a(p_1){\bf W}_\a(p_2)~){\rm Tr}(\bar{{\bf W}}^\ad(p_3)
\bar{{\bf W}}_\ad(p_4))\right.\nonumber\\
&&~~~~~~~~~~~~~~~~~~~~~~~~~~~~\left.
-\frac{1}{2} ~{\rm Tr}({\bf W}^\a(p_1)\bar{{\bf W}}^\ad(p_2))
~{\rm Tr}({\bf W}_\a(p_3)\bar{{\bf W}}_\ad(p_4))\right]
\label{PPTT}
\eea
 We introduce a mass IR
regulator, in order to avoid divergences which would arise in the zero limit
of the external $p_i$ momenta, and we perform the loop integration
using Schwinger parameters for
the denominator factors from the propagators.
In this way we obtain
\bea
I_0&=&\int d^4k~ \frac{e^{i(p_1+p_2)\times k}}
{[(k+p_1)^2+m^2][k^2+m^2][ (k-p_4)^2+m^2][(k+p_1+p_2)^2+m^2]}
\nonumber\\
&&~~~~\nonumber\\
&=&\int_0^\infty \prod_{i=1}^4 ~d\a_i~e^{-\a
m^2}~\int d^4k~
e^{-\a k^2}
~ e^{-ik\times (p_1+p_2)}~e^{\frac{i}{\a}[\a_1 p_1\times
p_2-\a_3 p_4\times (p_1+p_2)]}\nonumber\\
&&~~~~\nonumber\\
&&~~~~~~~~~~~~~~~~~~~
e^{\frac{1}{\a}
[-(\a_1+\a_4)(\a_2+\a_3)p_1^2-(\a_1+\a_2+\a_3)\a_4p_2^2-(\a_1+\a_2+\a_4)
\a_3p_4^2]}\nonumber\\
&&~~~~\nonumber\\
&&~~~~~~~~~~~~~~~~~~~~~~~~~~~~~~
e^{-\frac{2}{\a}[(\a_2+\a_3)\a_4p_1\cdot p_2
+(\a_1+\a_4)\a_3p_1\cdot p_4
+\a_3\a_4 p_2\cdot p_4]}\nonumber\\
&&~~~~\nonumber\\
&=&\int_0^\infty \prod_{i=1}^4
~d\a_i~\frac{1}{\a^2}~e^{-\frac{1}{\a}(p_1+p_2)\circ(p_1+p_2)}
e^{-\a m^2}~e^{\frac{i}{\a}[\a_1 p_1\times
p_2-\a_3 p_4\times (p_1+p_2)]}\nonumber\\
&&~~~~\nonumber\\
&&~~~~~~~~~~~~~~~~~~~~
e^{\frac{1}{\a}
[-(\a_1+\a_4)(\a_2+\a_3)p_1^2-(\a_1+\a_2+\a_3)\a_4p_2^2-(\a_1+\a_2+\a_4)
\a_3p_4^2]}\nonumber\\
&&~~~~\nonumber\\
&&~~~~~~~~~~~~~~~~~~~~~~~~~~~~~~~
e^{-\frac{2}{\a}[(\a_2+\a_3)\a_4p_1\cdot p_2
+(\a_1+\a_4)\a_3p_1\cdot p_4
+\a_3\a_4p_2\cdot p_4]}
\eea
where we have defined
\beq
p\circ p \equiv p_\m \Theta^2_{\m\n} p_\n
\eeq
and
\beq
 \a=\a_1+\a_2+\a_3+\a_4
\eeq
Introducing new integration variables
\beq
\l\equiv\a=\a_1+\a_2+\a_3+\a_4\qquad\qquad
\xi_i=\frac{\a_i}{\a}\qquad\qquad\qquad i=1,2,3,4
\label{newvar}
\eeq
we obtain
\bea
I_0&=&\int_0^\infty d\l~\l~e^{-\frac{1}{\l}(p_1+p_2)\circ(p_1+p_2)}
e^{-\l m^2}\int_0^1~ \prod_{i=1}^4
~d\xi_i~\d(1-\sum\xi_i)\nonumber\\
&&~~~~\nonumber\\
&&~~~~~~~
e^{i\xi_1 p_1\times
p_2-i\xi_3 p_3\times p_4}~e^{-\l[(\xi_1+\xi_4)(\xi_2+\xi_3)p_1^2+
(1-\xi_4)\xi_4
p_2^2+(1-\xi_3)
\xi_3p_4^2]}\nonumber\\
&&~~~~\nonumber\\
&&~~~~~~~~~~~~~~~~~
e^{-2\l[(\xi_2+\xi_3)\xi_4 p_1\cdot p_2
+(\xi_1+\xi_4)\xi_3p_1\cdot p_4
+\xi_3\xi_4 p_2\cdot p_4]}
\label{newint}
\eea
This result reproduces the field theory limit obtained from string
amplitudes in the presence of a nonzero $B$ field \cite{*trek,variouspeople}.
 The contributions from
all the other nonplanar supergraphs can be written in straightforward
manner.  A more detailed
presentation and a complete analysis will be given elsewhere \cite{SZ}.

In the low energy
approximation $p_i\cdot p_j$ small, with $p_i\times p_j$ finite, the
integration on the $\l$ and $\xi_i$ variables can be performed exactly
\cite{*trek}.
The final result from diagrams containing two $P$ and two $T$
vertices is given by \cite{SZ}
\bea
&&\G_{2P2T}\rightarrow\frac{1}{32}~\int d^2\theta~d^2\bar{\theta}~
\frac{d^4p_1~d^4p_2~d^4p_3~d^4p_4}{(2\p)^{16}}~\d(\sum
p_i)~
e^{-\frac{i}{2}p_1\times p_2}~e^{\frac{i}{2}p_3\times p_4} \nonumber\\
&&~~~~~~~~~~\nonumber\\
&&~~~~~~~~~~~~~~~~\frac{1}{2}\left[{\rm Tr}({\bf W}^\a(p_1){\bf W}_\a(p_2)~)
{\rm Tr}(\bar{{\bf W}}^\ad(p_3)
\bar{{\bf W}}_\ad(p_4))\right.\nonumber\\
&&~~~~~~~~~~~~~~~~~~~~~~~~~~
+ ~{\rm Tr}({\bf W}^\a(p_1)\bar{{\bf W}}^\ad(p_2))
~{\rm Tr}(\bar{{\bf W}}_\ad(p_3){\bf W}_\a(p_4))\nonumber\\
&&~~~~~~~~~~~~~~~~~~~~~~~~~~~~~~~~~\left.
- ~{\rm Tr}({\bf W}^\a(p_1)\bar{{\bf W}}^\ad(p_2))
~{\rm Tr}({\bf W}_\a(p_3)\bar{{\bf W}}_\ad(p_4))\right]\nonumber\\
&&~~~~~~~~~~~~~~~~~~\int_0^\infty d\l~\l~e^{-\frac{1}{\l}(p_1+p_2)\circ(p_1+p_2)}
e^{-\l m^2}\int_0^1
~d\xi_1~d\xi_3~e^{i\xi_1 p_1\times
p_2-i\xi_3 p_3\times p_4}\nonumber\\
&&~~~~\nonumber\\
&&=\frac{1}{2}~\int d^2\theta~d^2\bar{\theta}~
\frac{d^4p_1~d^4p_2~d^4p_3~d^4p_4}{(2\p)^{16}}~\d(\sum
p_i)~\frac{(p_1+p_2)\circ(p_1+p_2)}{2}\nonumber\\
&&~~~~\nonumber\\
&&~~~~~~~~~~~~~~~~\frac{\sin\left(\frac{p_1\times p_2}{2}\right)}
{\frac{p_1\times
p_2}{2}}~
\frac{\sin\left(\frac{p_3\times p_4}{2}\right)}{\frac{p_3\times
p_4}{2}}~
K_2(2m\sqrt{(p_1+p_2)\circ(p_1+p_2)}) \nonumber\\
&&~~~~~~~~~~\nonumber\\
&&~~~~~~~~~~~~~~~~\frac{1}{2}\left[{\rm Tr}({\bf W}^\a(p_1){\bf W}_\a(p_2)
)~{\rm Tr}(\bar{{\bf W}}^\ad(p_3)
\bar{{\bf W}}_\ad(p_4))\right.\nonumber\\
&&~~~~~~~~~~~~~~~~~~~~~~~~~~
+ ~{\rm Tr}({\bf W}^\a(p_1)\bar{{\bf W}}^\ad(p_2))
~{\rm Tr}(\bar{{\bf W}}_\ad(p_3){\bf W}_\a(p_4))\nonumber\\
&&~~~~~~~~~~~~~~~~~~~~~~~~~~~~~~~\left.
- ~{\rm Tr}({\bf W}^\a(p_1)\bar{{\bf W}}^\ad(p_2))
~{\rm Tr}({\bf W}_\a(p_3)\bar{{\bf W}}_\ad(p_4))\right]
\label{final}
\eea
where $K_2$ is the modified Bessel function.

The low-energy effective action contribution
in (\ref{final}) cannot be rewritten, as it was the case for the
planar diagrams, in terms of $*$-products of field strengths.
Moreover gauge invariance under the transformations in
(\ref{gaugetransf}) is not maintained. On the other hand
we are reassured that nothing went
wrong in the quantization procedure, since our
perturbative field theory result is in complete agreement with
the field theory limit from one-loop four-point scattering on
$D3$-branes as computed in \cite{*trek,gar}. It would be interesting
to see if one can implement gauge invariant
operators perturbatively, using techniques suggested in
\cite{GHI,IKK}.
In any event it appears that nonlocality, which is an intrinsic
property of
the noncommutative theory, might require a modified and deeper
understanding of the concept of gauge invariance.

\vspace{1.5cm}
\noindent
{\bf Acknowledgements}

\noindent
This work has been partially supported by INFN, MURST and the
European Commission TMR program ERBFMRX-CT96-0045, in which D.Z. is
associated to the University of Torino.

\end{document}